\documentclass[twocolumn]{revtex4}

\begin{document}

\title{Exact solutions of Brans-Dicke cosmology and the cosmic
coincidence problem}

\author{S. Carneiro$^{1,2}$}

\author{A. E. Montenegro Jr.$^1$}

\affiliation{$^1$Instituto de F\'{\i}sica, Universidade Federal da
Bahia, 40210-340, Salvador, BA, Brazil \\ $^2$International Centre
for Theoretical Physics, Trieste, Italy\footnote{Associate
Member}}

\begin{abstract}
We present some cosmological solutions of Brans-Dicke theory,
characterized by a decaying vacuum energy density and by a
constant relative matter density. With these features, they shed
light on the cosmological constant problems, leading to a
presently small vacuum term, and to a constant ratio between the
vacuum and matter energy densities. By fixing the only free
parameter of our solutions, we obtain cosmological parameters in
accordance with observations of the relative matter density, the
universe age and redshift-distance relations.
\end{abstract}

\maketitle

Several authors have been considering the possibility of a varying
cosmological term in order to fit the observed non-decelerated
expansion of the universe and, at the same time, to explain the
small value of the cosmological constant observed at present
\cite{Bertolami}-\cite{GRF2005}. Such a variation of the vacuum
density has found support in some quantum field approaches (see,
for example, \cite{Ralf,Shapiro}), in which context an induced
variation of the gravitational coupling constant $G$ may also be
expected \cite{Ademir2,MGX,Friedmann,GRF2005}.

Our goal in this contribution is to present some cosmological
solutions with decaying vacuum in the realm of Brans-Dicke theory.
For this purpose, we will consider an empirical variation law for
$G$, given by the Weinberg relation $G\approx H/m_\pi^3$, where
$H=\dot{a}/a$ is the Hubble parameter, and $m_\pi$ is the energy
scale of the QCD chiral symmetry breaking, the latest cosmological
vacuum phase transition. Such a relation, originally based on the
Eddington-Dirac large number coincidence, can find some support on
theoretical, holographic arguments \cite{Friedmann,Mena}. Let us
write it as
\begin{equation} \label{Weinberg}
G=\frac{H}{8\pi\lambda},
\end{equation}
where the constant $\lambda$ is positive and has the order of
$m_\pi^3$.

With this variation law for $G$, we will form an empirical ansatz
to be used in Brans-Dicke equations. It is fulfilled by the
additional constraint
\begin{equation} \label{rho}
\rho=\frac{3\alpha H^2}{8\pi G},
\end{equation}
where $\rho = \rho_m + \rho_{\Lambda}$ is the total energy
density, and $\alpha$ is a positive constant of the order of
unity. As we know, in the context of scalar-tensor theories, even
for zero spatial curvature, the total energy density is not
necessarily equal to the critical density $\rho_c\equiv3H^2/(8\pi
G)$. Therefore, the above equation should also be considered an
empirical coincidence, suggested by observation. Our point is
that, since (\ref{Weinberg}) and (\ref{rho}) are valid nowadays,
they may be valid for any time, or at least in the limit of late
times.

Let us find solutions for this limit, by considering a spatially
flat FLRW space-time, with a cosmic fluid formed by dust matter
(i.e., $p_m=0$) plus a vacuum term with equation of state
$p_\Lambda=-\rho_\Lambda$. The Brans-Dicke equations are then
given by \cite{BD,Weinberg}
\begin{equation} \label{1}
\frac{d(\dot{\phi}
a^3)}{dt}=\frac{8\pi}{3+2\omega}\left(\rho-3p\right)a^3=
\frac{8\pi}{3+2\omega}\left(\rho+3\rho_\Lambda\right)a^3,
\end{equation}
\begin{equation} \label{2}
\dot{\rho}=-3H(\rho+p)=-3H\rho_m,
\end{equation}
\begin{equation} \label{3}
H^2=\frac{8\pi\rho}{3\phi}-\frac{\dot\phi}{\phi}H+\frac{\omega}{6}\frac{\dot\phi^2}{\phi^2},
\end{equation}
where $p=p_m+p_{\Lambda}$ is the total pressure, $\omega$ is the
Brans-Dicke parameter, and the Brans-Dicke scalar field $\phi$ is
related to the gravitational constant by $\phi=G_0/G$, $G_0$ being
a positive constant of the order of unity.

From our ansatz (\ref{Weinberg})-(\ref{rho}), we obtain
\begin{equation}\label{4}
\rho=3\alpha\lambda H,
\end{equation}
\begin{equation}\label{5}
\phi=\frac{8\pi\lambda G_0}{H},
\end{equation}
\begin{equation}\label{6}
\dot{\phi}=8\pi\lambda G_0(1+q),
\end{equation}
where $q=-a\ddot{a}/\dot{a}^2$ is the deceleration parameter.

By using (\ref{4})-(\ref{6}), we can rewrite equations
(\ref{1})-(\ref{3}) in the form
\begin{equation}\label{7}
(3+2\omega)\lambda G_0[\dot{q}+3(1+q)H]=3\alpha\lambda
H+3\rho_\Lambda,
\end{equation}
\begin{equation}\label{8}
\rho_m=\alpha\lambda (1+q)H,
\end{equation}
\begin{equation}\label{9}
\frac{\alpha}{G_0}=2+q-\frac{\omega}{6}(1+q)^2.
\end{equation}
Equation (\ref{9}) shows that $q$ is a constant, and (\ref{7})
reduces to
\begin{equation}\label{10}
(3+2\omega)\lambda G_0(1+q)H=\alpha\lambda H+\rho_\Lambda.
\end{equation}

Using (\ref{4}) and (\ref{8}), we obtain a decaying vacuum
density, given by
\begin{equation}\label{11}
\rho_\Lambda=\alpha\lambda (2-q)H.
\end{equation}
Leading (\ref{Weinberg}) into (\ref{11}), and using
$\rho_\Lambda=\Lambda/8\pi G$, we see that the cosmological term
scales as
\begin{equation} \label{Lambda}
\Lambda=\alpha(2-q)H^2.
\end{equation}

Equations (\ref{11}) and (\ref{Lambda}) are curious results,
because they can also be derived on the basis of theoretical
reasoning \cite{Ademir,Ralf,Aldrovandi,Shapiro,Friedmann}, with no
use of the empirical relations we are considering here. We will
see that relations (\ref{11}) or (\ref{Lambda}) can be used to
form a second ansatz, with a larger set of solutions.

Substituting (\ref{4}) and (\ref{8}) into (\ref{2}), we obtain a
first order differential equation for $H$, which solution is given
by
\begin{equation}\label{12}
H = \left(\frac{1}{1+q}\right)\frac{1}{t}.
\end{equation}
Here, an integration constant was made zero in order to obtain the
divergence of $H$ at $t=0$. A second integration leads to a scale
factor evolving as
\begin{equation}\label{13}
a=At^{\frac{1}{1+q}},
\end{equation}
where $A$ is an arbitrary integration constant.

With the help of (\ref{Weinberg}), one can write the critical
density as $\rho_c=3\lambda H$. Therefore, from (\ref{8}) we
obtain a relative matter density
\begin{equation}\label{14}
\Omega_m\equiv\frac{\rho_m}{\rho_c}=\frac{\alpha(1+q)}{3}.
\end{equation}
This is perhaps the most interesting feature of the present
solutions. It means that, in the limit of late times we are
considering here, the relative matter density is a constant. This
is a consequence of the conservation of the total energy,
expressed by equation (\ref{2}). The vacuum decay is only possible
if associated to a process of matter production (a general feature
of vacuum states in non-stationary space-times). On the other
hand, since in our ansatz $\rho=\alpha\rho_c$ (see (\ref{rho})),
the constancy of $\Omega_m$ is a possible solution to the cosmic
coincidence problem, that is, the unexpected approximate
coincidence between the matter density and the dark energy
density. We will see below that (\ref{14}) is in accordance with
present observations.

Substituting $\rho_\Lambda$ from (\ref{11}) into (\ref{10}), one
has
\begin{equation}\label{17}
\frac{\alpha}{G_0}=\frac{(3+2\omega)(1+q)}{3-q}.
\end{equation}
Comparing the values of $\alpha/G_0$ given by (\ref{9}) and
(\ref{17}), we obtain a relation between $\omega$ and $q$:
\begin{equation}\label{18}
(3+2\omega)(1+q)=\left[2+q-\frac{\omega}{6}(1+q)^2\right](3-q).
\end{equation}
We can also, by eliminating $\omega$ from (\ref{9}) and
(\ref{17}), derive $\alpha/G_0$ as a function of $q$:
\begin{equation}\label{19}
\frac{\alpha}{G_0}=\frac{12(2+q)+3(1+q)^2}{(1+q)(3-q)+12}.
\end{equation}

With all these results we can estimate corresponding values for
the cosmological parameters. For example, for $q=0$ (that is, for
$a=At$) we obtain, from equation (\ref{18}), $\omega=6/5$; from
(\ref{19}) we have $\alpha/G_0=9/5$; from (\ref{12}) it follows
$Ht=1$; and from (\ref{14}) one obtains $\Omega_m=\alpha/3$.

As $\alpha\approx1$, we see that in this case $\Omega_m \approx
0.3$, a result corroborated by astronomical estimations
\cite{omega}. On the other hand, an age parameter $Ht\approx1$ has
been suggested by globular clusters observations \cite{age}.
Finally, a coasting expansion with $q\approx0$ is consistent with
observations of distance-redshift relations for supernova Ia and
compact radius sources \cite{Dev,Dev2,Alcaniz}.

As to the Brans-Dicke parameters $\omega$ and $G_0$ are concerned,
they are positive and of the order of unity. This could be
considered a bad result, in view of the high lower limits imposed
to $\omega$ by astronomical tests in the Solar System.
Nevertheless, let us remember that we are concerned here to a
vacuum density variation (and to a corresponding induced $G$
variation) at the cosmological scale, variations that cannot be
ruled out by observations at the local scale (where the metric is
stationary, and where any spatial dependence of $G$, related to
some spatial dependence of the vacuum density, is negligible). In
this sense, the Brans-Dicke theory we are using here, with
constant $\omega$, must be considered an effective description,
valid only in the cosmological limit \cite{GRF2005}. A more
general approach can be based on scalar-tensor theories in which
$\omega$ depends on the scale, being very high in the weak field
approximation of Solar System.

Another point to be considered is the constant character of the
deceleration parameter in our solutions. Although the present
observations indicate a non-decelerating expansion, an earlier
decelerated phase is usually expected in order to allow structure
formation. In spite of the claim of some authors \cite{Dev2} about
the possibility of a coasting expansion (i.e., with $q\approx0$)
even for earlier times, a more conservative viewpoint would be to
consider our empirical ansatz (\ref{Weinberg})-(\ref{rho}) valid
only in the limit of late times, as we have been done.

We can also replace our original ansatz by a more general one (in
the sense of presenting a larger set of solutions). We have seen
that our results (\ref{11}) and (\ref{Lambda}) (with constant $q$)
have been justified in different theoretical approaches to the
vacuum energy in curved space-times (for instance, in
\cite{Ralf,Shapiro}). We can therefore substitute $\Lambda=\beta
H^2$ (or, equivalently, $\rho_{\Lambda}=\beta \lambda H$, where
$\beta$ is a constant) for our empirical relation (\ref{rho}),
retaining the Weinberg relation (\ref{Weinberg}). As this last one
may also be understood on the basis of theoretical arguments (as
in \cite{Friedmann}, for example), this new ansatz should be
considered more justified from a theoretical viewpoint.

The set of solutions we can find leading the new ansatz into
Brans-Dicke equations (\ref{1})-(\ref{3}) will be shown in a
forthcoming publication. Let us just mention that we re-obtain, as
a particular case, the same solutions we have obtained with the
former ansatz. In addition, we have three other cosmological
solutions, for which the Brans-Dicke parameter is $\omega=-1$, and
the vacuum energy density is negative. In two of these additional
solutions, the deceleration parameter is always highly positive or
always highly negative, being them therefore ruled out by
observations.

The third new solution has an initial singularity, an early
decelerating phase followed by an accelerated one, and a
``big-rip" singularity, with $a$, $H$ and $\rho_m$ diverging at a
finite time in the future (but with $\Omega_m$ remaining finite).
So, it is an interesting solution from a theoretical perspective.
Unfortunately, for $q$ in the range given by the present
observations ($-1<q<1$), the age parameter for this solution is
less than $Ht\approx0.5$, outside the observed limits.

${ }$

S.C. is grateful to E. Abdalla, R. Abramo, A. Saa, and the other
organizers of ``100 Years of Relativity", for the support and warm
hospitality in S\~ao Paulo.

\end{document}